\documentclass[final]{svjour3}
\usepackage{amsmath}
\usepackage{graphicx}
\usepackage{color}
\usepackage{url}

\def\makeheadbox{{%
\hbox to0pt{\vbox{\baselineskip=10dd\hrule\hbox
to\hsize{\vrule\kern3pt\vbox{\kern3pt
\hbox{\bfseries Paper submitted to Journal of Low Temperature Physics}
\kern3pt}\hfil\kern3pt\vrule}\hrule}%
\hss}}}

\begin{document}

\title{Gauge invariance of thermal transport coefficients}

\author{Loris Ercole$^\mathbf{1}$ \and Aris Marcolongo$^\mathbf{2}$
  \and \\ Paolo Umari$^\mathbf{3}$ \and 
  Stefano Baroni$^\mathbf{1}$ }
\authorrunning{Loris Ercole et al.}

\dedication{This paper is dedicated to our friend and distinguished colleague
Flavio Toigo, a long-time enthusiast of hydrodynamic fluctuations,
on the occasion of his seventieth birthday.}

\institute{
  \begin{enumerate}
  \item SISSA -- Scuola Internazionale Superiore di Studi Avanzati, Via
  Bonomea 265, 34136 Trieste, Italy
  \item Theory and Simulation of Materials (THEOS), \'Ecole Polytechnique F\'ed\'erale
  de Lausanne, 1015 Lausanne, Switzerland
  \item  Dipartimento di Fisica ed Astronomia, Universit\`a di Padova, via Marzolo
  8, 35131 Padova, Italy 
\end{enumerate}
}

\date{November 30, 2015}
\maketitle

\begin{abstract}
  Thermal transport coefficients are independent of the specific
  microscopic expression for the energy density and current from which
  they can be derived through the Green-Kubo formula. We discuss
  this independence in terms of a kind of \emph{gauge invariance
  }resulting from energy conservation and extensivity, and demonstrate
  it numerically for a Lennard-Jones fluid, where different forms of
  the microscopic energy density lead to different time correlation
  functions for the heat flux, all of them, however, resulting in the
  same value for the thermal conductivity.
\end{abstract}

\newpage It has long been thought that the inherent indeterminacy of
any quantum mechanical expression for the energy density would hinder
the evaluation of thermal transport coefficients from equilibrium
\emph{ab-initio} molecular dynamics (AIMD), using the Green-Kubo (GK)
formalism \cite{Green:1954,Kubo:1957,Kadanoff:1963,Forster}.  In
classical molecular dynamics (CMD) this goal is achieved by
decomposing the total energy of an extended system into localized
atomic contributions and by deriving from this decomposition an
explicit (and allegedly unique) expression for the energy flux. While
the calculation of thermal transport coefficients from equilibrium
AIMD has been successfully addressed by some of us in a recent work
\cite{Marcolongo:2015}, the question still remains as to whether the
expression for the energy flux currently used in CMD is uniquely
defined and, in the negative, how is it that different definitions of
the energy flux would lead to the same value for the thermal
conductivity. In this paper we show that different equivalent
definitions for the atomic energies in a classical system lead to
different expressions for the macroscopic energy flux, and demonstrate
numerically in the case of a Lennard-Jones fluid that these
expressions result in the same value for the thermal conductivity, as
evaluated from equilibrium CMD through the GK formula.  This finding
is then rationalized in terms of a kind of \emph{gauge invariance} of
heat transport coefficients, resulting from energy conservation and
extensivity.

According to the GK formalism
\cite{Green:1954,Kubo:1957,Kadanoff:1963,Forster}, the heat
conductivity $\kappa$ of an isotropic material can be expressed in
terms of the auto-correlation function of the macroscopic heat flux,
$\mathbf{J}_{q}(t)$, as:
\begin{equation}
  \kappa=\frac{1}{3Vk_{B}T^{2}}\int_{0}^{\infty}\langle
  \mathbf{J}_{q}(t) \cdot\mathbf{J}_{q}(0)\rangle
  dt,\label{eq:Green-Kubo} 
\end{equation}
where brackets $\langle\cdot\rangle$ indicate canonical averages,
$k_{B}$ is the Boltzmann constant, and $V$ and $T$ are the system
volume and temperature, respectively. The heat flux is the macroscopic
average of the heat current density, which is in turn defined as the
non-convective component of the energy current density. Atoms in
solids can be assumed to not diffuse, while in one-component and
molecular fluids convective energy transport can
be disregarded because of momentum conservation. Because of this, in
the following we assume that energy and heat currents coincide.

\begin{figure}
\begin{minipage}{0.5\textwidth}
\centering \includegraphics{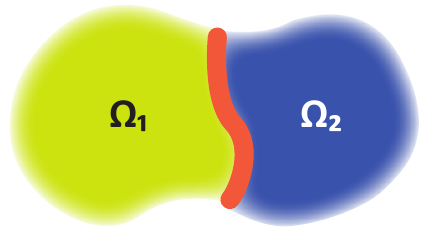}
\end{minipage}
\begin{minipage}{0.4\textwidth}
\begin{align*}
  E(\Omega_{1}\cup\Omega_2) &= E(\Omega_1) + E(\Omega_2) + W_{12}
                              \qquad \\
  & \overset{?}{=}\mathcal{E}(\Omega_1)+\mathcal{E}(\Omega_2)
\end{align*}
\end{minipage}
\caption{
  The energy of an isolated system is the sum of the energies of its
  subsystems (as defined when they are isolated as well) plus the
  interaction among them, $W$, whose magnitude scales as the area of the
  interface, depicted in red. When defining the energies of individual
  subsystems, $\mathcal{E}$, $W$ has to be
  arbitrarily partitioned among them. 
  \label{fig:energy-partition}
}
\end{figure}

In CMD the macroscopic energy flux is expressed in terms of suitably
defined atomic energies whose sum yields the total energy of the
system.  For the sake of simplicity, we restrict our attention to
one-component systems held together by pair potentials, in which case
the atomic energies can be defined as \cite{Hansen:2006}:
\begin{equation}
  \epsilon_{I}(\mathbf{R},\mathbf{V}) = \frac{1}{2}MV_{I}^{2} +
  \frac{1}{2}\sum_{J\ne I}
  v(|\mathbf{R}_{I}-\mathbf{R}_{J}|), \label{eq:atomic-energies} 
\end{equation}
where $M$ is the atomic mass, $\mathbf{R}\doteq\{\mathbf{R}_{I}\}$ and
$\mathbf{V}\doteq\{\mathbf{V}_{I}\}$ are atomic coordinates and
velocities, respectively, $v(R)$ is the inter-atomic pair potential,
and the indices $I$ and $J$ run over all the atoms in the system.
Using standard manipulations \cite{Hansen:2006,Lee:1991}, the macroscopic
energy flux can be obtained from Eq. \eqref{eq:atomic-energies} as:
\begin{equation}
  \mathbf{J}_{e}=\sum_{I}\epsilon_{I}\mathbf{V}_{I} - \frac{1}{2}
  \sum_{I, J\ne I}\bigl (\mathbf{V}_{I}\cdot \nabla_Iv(|\mathbf{R}_I -
  \mathbf{R}_J|) \bigr )
  (\mathbf{R}_{I}-\mathbf{R}_{J}). \label{eq:classical-current-2potentials}
\end{equation}
where (for a derivation of
Eq. \eqref{eq:classical-current-2potentials}, valid for general
many-body inter-atomic potentials \cite{Lee:1991} and its
specialization to the present pair-potentials case, see the
Appendix). It is often implicitly assumed that the well-definedness of
thermal transport coefficients would stem from the uniqueness of the
decomposition of the system's total energy into localized, atomic,
contributions. This assumption is manifestly incorrect, as any
decomposition leading to the same value for the total energy as
Eq. \eqref{eq:atomic-energies} should be considered as legitimate. The
difficulty of partitioning a system's energy into subsystems'
contributions is illustrated in Figure \ref{fig:energy-partition}, which
depicts a system made of two interacting subsystems. When defining the
energy of each of the two susbsystems, an arbitrary decision has to be
made as to how the interaction energy is partitioned. In the case
depicted in Figure \ref{fig:energy-partition}, for instance, the energy
of each of the two subsystems can be defined as
$\mathcal{E}(\Omega_i) = E(\Omega_i) + \frac{1}{2}(1\pm\gamma)W_{12}$,
where $E(\Omega_i)$ are the energies of the two isolated subsystems,
$W_{12}$ their interaction energy, and $\gamma$ an arbitrary
constant. In the thermodynamic limit, when the energy of any relevant
subsystem is much larger than the interaction between any pairs of
them, the value of the $\gamma$ constant is irrelevant. When it comes
to definining energy densities (\emph{i.e.} energies of infinitesimal
portions of a system) or atomic energies, instead, the magnitude of
the interaction between different susbsystems is comparable to the
their energies, which become therefore intrinsically ill-defined.

As a specific example, we consider the
following definition for the atomic energies \cite{Marcolongo:2014}:
\begin{equation}
  \epsilon_{I}^{\Gamma}(\mathbf{R},\mathbf{V}) =
  \frac{1}{2}M_{I}\mathrm{V}_{I}^{2} + \frac{1}{2}\sum_{J\ne I}
  v(|\mathbf{R}_{I}-\mathbf{R}_{J}|)
  (1+\Gamma_{IJ}), \label{eq:gamma-atomic-energies} 
\end{equation}
where $\Gamma_{IJ}=-\Gamma_{JI}$ is \emph{any} antisymmetric matrix.
As the inter-atomic potential appearing in
Eq. \eqref{eq:gamma-atomic-energies} is symmetric with respect to the
atomic indices, it is clear that the sum of all the atomic energies
does not depend on $\Gamma$, thus making any choice of $\Gamma$
equally permissible. This trivial observation has deep consequences on
the theory of thermal fluctuations and transport, because the value of
the macroscopic energy flux, instead, depends explicitly on $\Gamma$,
thus making one fear that the resulting transport coefficients would
depend on $\Gamma$ as well. Using the same manipulations that lead
from Eq. \eqref{eq:atomic-energies} to
Eq. \eqref{eq:classical-current-2potentials}, for any choice of the $\Gamma$
matrix in Eq. \eqref{eq:gamma-atomic-energies}, a corresponding
expression for the macroscopic energy flux can be found, reading
\cite{Marcolongo:2014}:
\begin{equation}
\mathbf{J}_{e}^{\Gamma}=\mathbf{J}_{e}+\frac{1}{2}\sum_{I, J\ne
  I}\Gamma_{IJ} \Bigl ( v(|\mathbf{R}_{I} -
  \mathbf{R}_{J}|) \mathbf{V}_{I} + 
(\mathbf{V}_{I}\cdot  \nabla_Iv(|\mathbf{R}_I -
  \mathbf{R}_J|) )
  (\mathbf{R}_{I}-\mathbf{R}_{J}) \Bigr ). \label{eq:gamma-classical-current} 
\end{equation}

\begin{figure}
\centering \includegraphics{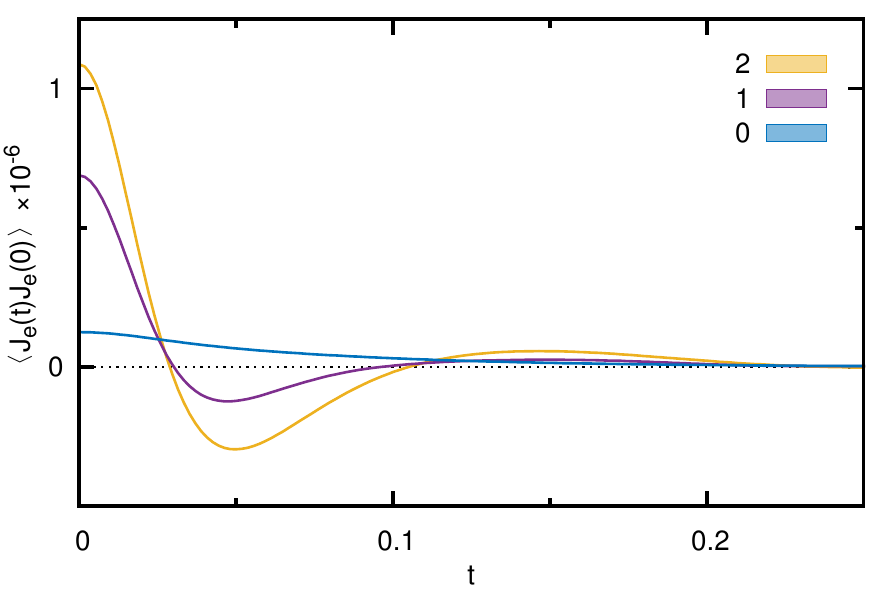}
\caption{Time correlation functions of the modified macroscopic energy
  flux, as defined in Eq. \eqref{eq:gamma-classical-current}, for
  different definitions of the $\Gamma$ matrix (see text). 
  The ``0'' line refers to the original definition (\emph{i.e.}
  $\Gamma = 0$), whereas the label ``1'' and ``2'' correspond to the
  two definitions of the $\Gamma$ matrix given in Eq. \eqref{eq:Gamma}.
  The parameters used are $\gamma_1 = 10$ and $\gamma_2 = 2.5$. 
  Error bars, as estimated by standard block analysis, are smaller 
  than the thickness of the lines. Units are Lennard-Jones units 
  ($M=\sigma=\varepsilon=1$). \label{fig:J0Jt} 
}
\end{figure}

\begin{figure}
\centering
\includegraphics{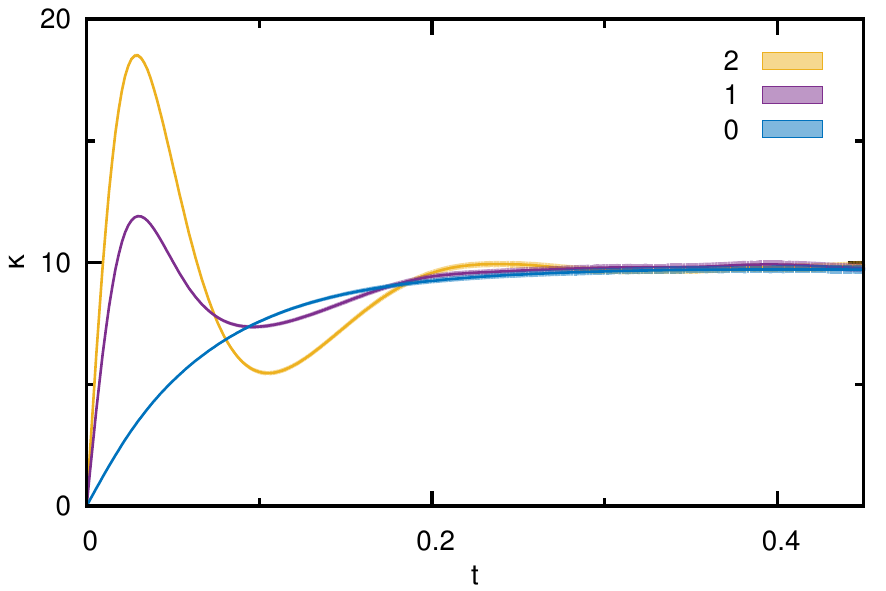}
\caption{Integral of the time correlation functions displayed in
  Figure \ref{fig:J0Jt}, multiplied by the prefactor appearing in the
  GK relation, Eq. \eqref{eq:Green-Kubo}, as a function of the upper
  limit of integration. Units are Lennard-Jones units (see caption to
  Figure \ref{fig:J0Jt}). The barely visible shaded area surrounding
  each line is an indication of the error bars, as estimated by
  standard block analysis. \label{fig:lambda}}
\end{figure}

In order to illustrate this state of affairs, we have performed CMD
simulations for a fluid of identical atoms, interacting through a
Lennard-Jones potential:
$v(R) = 4\varepsilon \left[ \left( \frac{\sigma}{R}\right)^{12} -
  \left(\frac{\sigma}{R} \right)^{6} \right]$
at density-temperature conditions $\rho=0.925\sigma^{-3}$ and
$T=1.86\varepsilon/k_{B}$, using cubic simulation cells containing 256
atoms in the iso-choric microcanonical ensemble, $(NVE)$
\cite{LAMMPS:1995}. In Figure \ref{fig:J0Jt} we display the macroscopic energy-flux
autocorrelation function corresponding to different choices of the
$\Gamma$ matrix in Eqs. \eqref{eq:gamma-atomic-energies} and
\eqref{eq:gamma-classical-current}. 
The $\Gamma$ matrices have been constructed in two different ways,
according to the prescriptions:
\begin{equation}
\setbox0=\vbox{\hsize=50mm \small\noindent where the matrix
      elements of $A$ are drawn from a uniform deviate  in the
      $[0,\gamma]$ interval.}
\setbox1 \vbox{\hsize=50mm \small\noindent according to whether $I=J$,
  $I>J$, or $I<J$.}
\Gamma_{IJ}= \left \{
  \begin{alignedat}{2}
    \frac{1}{2} \left(A_{IJ}-A_{JI}\right ) & \quad\raise -.5\ht0
    \box0 && \qquad (1) \\
\\
    0, +\gamma, -\gamma & \quad\raise -0.5\ht1 \box1 && \qquad (2) \\
  \end{alignedat} \right .
\label{eq:Gamma}
\end{equation}
Figure \ref{fig:J0Jt} clearly shows that the
$\langle \mathbf{J}_{e}^{\Gamma}(t) \cdot\mathbf{J}_{e}^{\Gamma}(0)
\rangle$
correlation functions dramatically depend on the $\Gamma$ matrices in
Eqs. \eqref{eq:gamma-atomic-energies} and
\eqref{eq:gamma-classical-current}.  Notwithstanding, the integrals of
all these time correlation functions tend to the same limit at large
integration times, as displayed in Figure \ref{fig:lambda}.

In order to get insight into this remarkable invariance property, let
us inspect the difference between the generalized flux in Eq.
\eqref{eq:gamma-classical-current} and the standard expression of
Eq. \eqref{eq:classical-current-2potentials}:
\begin{equation}
  \Delta\mathbf{J}_{e}^{\Gamma} =\mathbf{J}_{e}^{\Gamma}-\mathbf{J}_{e}
  =\frac{\mathrm{d}}{\mathrm{dt}}\frac{1}{4} \sum_{I, J\ne I}
    \Gamma_{IJ} \, v(|\mathbf{R}_{I}-\mathbf{R}_{J}|)
    (\mathbf{R}_{I}-\mathbf{R}_{J}). \label{eq:DeltaJ}
\end{equation}
We see that the two different expressions for the macroscopic energy
flux differ by a total time derivative. In the following, we will show
that this is a consequence of energy conservation and extensivity and
a sufficient condition for the corresponding thermal conductivity to coincide.

The very possibility of defining an energy current density stems from
energy extensivity and conservation. Energy is extensive: because of
this, the energy of a macroscopic sample of matter of volume $\Omega$
can be written as the integral of an energy density, $e(\mathbf{r})$:
\begin{equation}
  E[\Omega]=\int_{\Omega}e(\mathbf{r})d\mathbf{r}. \label{eq:extensivity}
\end{equation}
Of course, the energy density appearing in Eq. \eqref{eq:extensivity}
is not uniquely defined, the only requirement being that its integral
over a domain $\Omega$ is well defined in the thermodynamic limit,
\emph{i.e.} two different densities whose integral over a domain
$\Omega$ differ by a quantity that scales as the area of the domain
boundary should be considered as equivalent. This equivalence can be
expressed by the condition that two equivalent densities, say
$e_{1}(\mathbf{r})$ and $e_{2}(\mathbf{r})$, differ by the divergence
of a (bounded) vector field:
\begin{equation}
  e_{2}(\mathbf{r})=e_{1}(\mathbf{r})+\nabla\cdot
  \mathbf{p}(\mathbf{r}). \label{eq:gauge_transformation}
\end{equation}
In a sense, two equivalent energy densities can be thought of as
different \emph{gauges }of the same scalar field.

Energy is also conserved: because of this, for any given gauge of
the energy density, $e(\mathbf{r})$, an energy current density can
be defined, $\mathbf{j}_{e}(\mathbf{r},t)$, so as to satisfy the
continuity equation:
\begin{equation}
  \dot{e}(\mathbf{r},t)=-\nabla\cdot
  \mathbf{j}_{e}(\mathbf{r},t), \label{eq:continuity} 
\end{equation}
where the dot indicates a time derivative. By combining
Eqs. \eqref{eq:gauge_transformation} and \eqref{eq:continuity} we see
that energy current densities and macroscopic fluxes transform under a
gauge transformation as:
\begin{align}
  \mathbf{j}_{2}(\mathbf{r},t) & = \mathbf{j}_{1}(\mathbf{r},t)-
                                 \dot{\mathbf{p}}(\mathbf{r},t), \label{eq:current_density_gauge}
  \\   \mathbf{J}_{2}(t) & =
                           \mathbf{J}_{1}(t)-\dot{\mathbf{P}}(t), \label{eq:macroscopic_flux_gauge} 
\end{align}
where $\mathbf{P}(t)=\int\mathbf{p}(\mathbf{r},t)d\mathbf{r}$. We
conclude that the macroscopic energy fluxes in two different energy
gauges differ by the total time derivative of a vector.

Our previous findings on the energy flux of a system of classical
atoms interacting through pair potentials as embodied in
Eq. \eqref{eq:DeltaJ} can be recovered by defining the corresponding
energy density as:
\begin{equation}
  e(\mathbf{r})=\sum_I \delta(\mathbf{r}-\mathbf{R}_I)
  \epsilon_I \label{eq:classical-density}.
\end{equation} 
By taking the first moment of the continuity equation,
Eq. \eqref{eq:continuity}, with respect to $\mathbf{r}$ and
integrating by parts its right-hand side, one sees that the
macroscopic average of the energy current density is the first moment
of the time derivative of the energy density:
\begin{equation}
  \mathbf{J}_e(t)=\int \dot e(\mathbf{r},t) \mathbf{r}
  ~d\mathbf{r}. \label{eq:first-moment} 
\end{equation}
Eq. \eqref{eq:first-moment} is ill-defined in periodic boundary
conditions because the position variable appearing therein is defined
modulo an integer translation, and the first moment of a periodic
function depends therefore on the definition and choice of origin of
the unit cell. This same difficulty affects the definition of the
macroscopic polarization in insulators and has given birth to the so
called \emph{modern theory of polarization} \cite{Resta:2007}.
Nevertheless, by throwing Eq. \eqref{eq:classical-density} into
Eq. \eqref{eq:first-moment} and using Newton's equations of motion,
Eq. \eqref{eq:first-moment} can be cast into a boundary-insensitive
form, as explained in detail in the Appendix, eventually resulting in the
expressions for the macroscopic energy flux given by Eqs.
\eqref{eq:classical-current} and \eqref{eq:gamma-classical-current}.

We now show that the energy fluxes of the same system in two different
energy gauges, $e_{1}$ and $e_{2}$, thus differing by a total time
derivative, as in Eq. \eqref{eq:macroscopic_flux_gauge}, result in the
same heat conductivity, as given by the Green-Kubo formula,
Eq. \eqref{eq:Green-Kubo}. Let us indicate by $\kappa_{1}$ and
$\kappa_{2}$ the thermal conductivities in the two gauges.  Using
Eq. \eqref{eq:macroscopic_flux_gauge} and the property that classical
time auto-correlation functions are even in time, one obtains:
\begin{multline}
  \kappa_{2} = \kappa_{1} + \\ \frac{1}{6Vk_{B}T^{2}}
  \int_{-\infty}^{+\infty}\frac{d}{dt}
  \bigl(\langle\textbf{P}(-t) \cdot\mathbf{J}_{1}(0) \rangle-
  \langle\textbf{P}(t) \cdot\mathbf{J}_{1}(0)\rangle+
  \langle\textbf{P}(t)\cdot
  \dot{\mathbf{P}}(0)
  \rangle\bigr)
  dt. \label{eq:kappa1_kappa2}
\end{multline}
The integral on the right-hand side of Eq. \eqref{eq:kappa1_kappa2}
vanishes because the correlation function of two observables at large
time lags factorizes into the product of two time-independent average
values and because the average value of the current $\mathbf{J}_{1}$, as
well as of any total time derivative, vanishes at equilibrium. We
conclude that the heat conductivities computed in different energy
gauges coincide, as they must on physical grounds.

In this paper we have demonstrated that, while the heat flux is
inherently undetermined at the atomic level, the heat conductivity
resulting from it through the Green-Kubo formula is indeed well
defined, as any measurable property must be. This indeterminacy stems
from the liberty one has to formally unpack the total energy of an
extended system into localized contributions in an infinite number of
equivalent ways.  We believe that this freedom can be exploited to
design the definition of the local energy (be it in terms of atomic
energies or energy densities), so as to optimise the convergence of
computer simulations, regarding simulation length, system size, or
both.

\section*{Appendix}
In order to derive Eq. \eqref{eq:classical-current} from
Eq. \eqref{eq:first-moment}, let us first compute the time derivative
of Eq. \eqref{eq:classical-density} using the definition of the
atomic energies, Eq. \eqref{eq:atomic-energies}, to obtain:
\begin{equation}
  \dot{e}(\mathbf{r}) =
  \sum_{I} \left[ \epsilon_{I} \mathbf{V}_{I}
    \cdot\nabla\delta(\mathbf{R}_{I}-\mathbf{r})
    + \delta(\mathbf{r}-\mathbf{R}_{I}) 
    \sum_{J}
    \left(
      \frac{\partial\epsilon_{I}}{\partial\mathbf{V}_{J}} \cdot
      \dot{\mathbf{V}}_{J}
      +\frac{\partial\epsilon_{I}}{\partial\mathbf{R}_{J}}
      \cdot\mathbf{V}_{J} 
    \right)
  \right],
  \label{eq:e-dot}
\end{equation}
where $\nabla\delta$ is the gradient of the $\delta$ function.
We now insert Eq. \eqref{eq:e-dot} into Eq. \eqref{eq:first-moment},
to obtain:
\begin{equation}
  \mathbf{J} = \sum_{I} \epsilon_{I} \mathbf{V}_{I} +
  \sum_{I}\mathbf{R}_{I} 
  \left(
    \mathbf{V}_{I}\cdot\mathbf{F}_{I}-\sum_{J} \mathbf{F}_{JI} \cdot
    \mathbf{V}_{J}
  \right), 
  \label{eq:J-A1}
\end{equation}
where $\mathbf{F}_{I}=M_{I}\dot{\mathbf{V}}_{I}$ is the force acting
on the $I$-th atom and
$\mathbf{F}_{JI}=-\frac{\partial
  \epsilon_{I}}{\partial\mathbf{R}_{J}}$.
Eq. \eqref{eq:J-A1} is not well defined in periodic boundary
conditions because the position of the $I$-th atom, $\mathbf{R}_I$, is
defined modulo a vector in the Bravais lattice and depends on the
choice of the origin of the unit cell.  In order to write a well
defined expression for the current, we first note that
$\mathbf{F}_{I} = \sum_{J}\mathbf{F}_{IJ}$. Inserting this relation
into Eq. \eqref{eq:J-A1}, one obtains:
\begin{equation}
  \mathbf{J}=\sum_{I} \epsilon_{I} \mathbf{V}_{I}+ \sum_{IJ}
  \mathbf{R}_{I}(\mathbf{V}_{I} \cdot \mathbf{F}_{IJ}-\mathbf{V}_{J}
  \cdot\mathbf{F}_{JI}). \label{eq:J_appendix}
\end{equation}
Now in the second term of the second sum in Eq. \eqref{eq:J_appendix}
we can interchange the $IJ$ dummy indeces, to obtain:
\begin{equation}
  \mathbf{J}=\sum_{I}\epsilon_{I}\mathbf{V}_{I} +
  \sum_{I, J\ne I}(\mathbf{V}_{I}\cdot\mathbf{F}_{IJ})
  (\mathbf{R}_{I}-\mathbf{R}_{J}). \label{eq:classical-current}
\end{equation}
Eq. \eqref{eq:classical-current} is well defined in periodic boundary
conditions, because it depends on atomic positions only through
differences between pairs of them. Note that this derivation does not
depend on the assumption that atoms interact through pair potentials,
and it holds in fact for general many-body inter-atomic
potentials. When atoms interact through pair potentials, as it is the
case considered in the present paper, and adopting the standard
definition for the atomic energies, Eq. \eqref{eq:atomic-energies},
one has:
$\mathbf{F}_{IJ} = \frac{1}{2}\bigl ( \delta_{IJ} \mathbf{F}_I -
\nabla_I v(|\mathbf{R}_I - \mathbf{R}_J|) \bigr )$,
where $\nabla_I$ indicates the gradient with respect to the position
of the $I$-th atom. Throwing this expression into
Eq. \eqref{eq:classical-current}, one one finally obtains Eq.
\eqref{eq:classical-current-2potentials} used in the text.  Similarly,
one would obtain the modified energy flux of
Eq. \eqref{eq:gamma-classical-current} from the definition for the
atomic energies given in Eq. \eqref{eq:gamma-atomic-energies}.


\begin{thebibliography}{10}

\bibitem{Green:1954}M.\,S. Green, J. Chem. Phys. \textbf{22}, 398
(1954).

\bibitem{Kubo:1957}R. Kubo, J. Phys. Soc. Jpn. \textbf{12}, 570 (1957).

\bibitem{Kadanoff:1963} L.\,P. Kadanoff and P.\,C. Martin, Ann.
Phys. \textbf{24}, 419 (1963).

\bibitem{Forster}D. Forster, \emph{Hydrodynamic fluctuations, broken
symmetry, and correlation functions} (Benjamin, Reading, 1975).

\bibitem{Marcolongo:2015} A. Marcolongo, P. Umari, and S. Baroni,
Nature Physics, doi:10.1038/nphys3509 (2015).

\bibitem{Hansen:2006} J.-P. Hansen and I.R. McDonald, \emph{Theory
of Simple Liquids} (Third Edition, Elsevier, Philadelphia, 2006).

\bibitem{Lee:1991} Y. Lee, R. Biswas, C. Soukoulis, C. Wang, C. Chan,
  and K.M. Ho, Phys. Rev. B, \textbf{43}, 6573 (1991).

\bibitem{Marcolongo:2014} A. Marcolongo, SISSA PhD thesis,
  \url{http://cm.sissa.it/thesis/2014/marcolongo}. 

\bibitem{LAMMPS:1995}CMD simulations have been performed using the
LAMMPS code, see: S. Plimpton, J. Comp. Phys. \textbf{117}, 1 (1995);
\url{http://lammps.sandia.gov}. 

\bibitem{Resta:2007} R. Resta and D. Vanderbilt,
  Top. Appl. Phys. \textbf{105}, 31 (2007).

\end{thebibliography}
\end{document}